# Resolving the Sodiation Process in Hard Carbon Anodes with Nanostructure Specific X-ray Imaging

Martina Olsson[1], Antoine Klein[1], Nataliia Mozhzhukhina[1,2], Shizhao Xiong[1], Christian Appel[3], Mads Carlsen[3], Leonard Nielsen[1], Linnea Rensmo[3,4], Marianne Liebi[1,3,4*], Aleksandar Matic[1*]

*1. Department of Physics, Chalmers University of Technology Gothenburg 41296, Sweden.*

*2. SEEL Swedish Electric Transport Laboratory, Säve Flygplatsväg 27, 42373 Säve, Sweden*

*3. Center for Photon Science, Paul Scherrer Institut, 5232 Villigen PSI, Switzerland*

*4. Institute of Materials, Ecole Polytechnique Fédérale de Lausanne (EPFL), 1015 Lausanne, Switzerland*

## Abstract

Hard carbons show significant promise as anode materials for sodium-ion batteries. However, monitoring the sodiation process in the hard carbon electrode during cycling and understanding the sodiation mechanism remain challenging. This article reports on *operando* 2D scanning small- and wide-angle X-ray scattering (SWAXS) and *ex situ* 3D SAXS tomography of hard carbon electrodes during the sodiation process. Structural changes are monitored with spatial and temporal resolution during the electrochemical process and shows that sodiation through micropore filling is the more dominating mechanism in the later stages of sodiation, i.e. in the plateau region of the voltage profile, while intercalation occurs continuously. Spatial inhomogeneities are resolved over the electrode and reveal an increased level of inhomogeneity at higher degree of sodiation with regions of different degrees of micropore filling. Resolving the processes spatially shows that plating starts at the interface between the electrode and the current collector where also a high degree of micropore filling and formation of pseudo-metallic sodium is found. The work demonstrates how SWAXS imaging can contribute to understanding the sodiation of hard carbon anodes, not only by spatially resolved analysis, but as a method to decouple contributions from different components in a cell, enabling more accurate scattering analysis in *in situ* environments.

# 1. Introduction

Sodium-ion batteries present a sustainable alternative to lithium-ion batteries due to the significantly higher abundance of sodium compared to lithium[1, 2]. The sodium-ion battery is in many respects analogous to the lithium-ion battery in terms of configuration of the electrodes and electrolytes, simplifying the implementation and development towards the market. However, the graphite electrodes used in lithium-ion batteries are not suitable for sodium-ion batteries as sodium does not intercalate into graphite at ambient pressure [3, 4]. Hard carbon, a disordered carbonaceous material, is instead considered as the primary anode material [5, 6]. The structure of hard carbon consists of stacked graphene layers arranged in a non-uniform, chaotic manner forming microporous, granular particles. The precise structure of the hard carbon material will depend on both the precursor and the pyrolysis conditions, where in general a higher pyrolysis temperature results in a larger microporosity and decreased spacing between stacked graphene layers as well as decreased defect concentrations in the hard carbon matrix.[7] This is reflected in a high diversity of structural features within the material class of hard carbons which contributes to the complexity to disentangle different sodiation mechanisms and a need for innovative characterization methods[8].

A typical voltage profile of sodiation of hard carbon entails a sloping region at higher potentials (≈0.1–2.5 V vs $Na/Na^+$) followed by a plateau region at lower potentials (≈0–0.1 V vs $Na/Na^+$), Figure 1a. It has been proposed that sodiation in hard carbon occurs by several different mechanisms: edge and defect adsorption, intercalation, and micropore filling[9-11]. However which mechanism determines the kinetics of sodiation in hard carbon, and which process is related to the different features in the electrochemical response, remains a question of controversy[12]. However, there is an increasing preference that edge and defect adsorption and intercalation are dominating in the sloping region and micropore filling occurs in the plateau region [8, 13-17]. In addition, unwanted sodium metal plating can occur in the hard carbon anode in the later stages of the electrochemical process which can lead to reduced capacity and cycle stability[18]. Due to the plateau region having a potential close to the one of sodium plating, this can be a serious concern particularly at high current rates[19].

Small- and wide-angle X-ray scattering (SWAXS) have been extensively used to characterize carbonaceous materials as it provides insight into its nanostructure and molecular arrangement [3, 13, 15, 20, 21]. It is particularly useful for studying the hierarchical structure of hard carbon, as it simultaneously probes the micropores (i.e. pores smaller than 2 nm) with SAXS, as well as the atomic correlation of graphene layers, accessed by WAXS. In this way, SWAXS has been used to characterize sodium insertion as the structural changes in the material revealed by the scattering signal can be related to a particular sodiation process and thereby reflect the sodiation mechanism [15, 16, 20]. Morikawa et al.[20] applied *ex situ* SWAXS studies to correlate the change in scattering intensity to sodium storage within both micropores and between graphene-graphene interlayers. Their study revealed that the storage mechanism of hard carbons depends on the microstructure and availability of storage sites between different carbons. In addition, they revealed a presence of pseudo-metallic sodium in the later stages of sodiation and related it to densely confined sodium in the micropores. Furthermore, SWAXS is a useful method for *in situ* and *operando* studies of hard carbon, as shown by Stevens and Dahn [16] and Iglesias et al.[15], enabling to directly monitor changes in the electrode structure during cycling. Stevens and Dahn correlated a decrease in scattering intensity from micropores to pore filling occurring at the low voltage plateau of the voltage profile. Iglesias et al. made a quantitative assessment of the pore filling process, showing that different pore sizes are preferentially filled during the sodiation process in different hard carbons.

Following these temporally resolved SWAXS studies, we here apply SWAXS imaging to study the sodiation of hard carbon with both temporal and spatial resolution to provide a new perspective for understanding the sodiation process. SWAXS imaging enables spatially resolved monitoring of structural changes during cycling which has recently been shown to reveal deformations in the interior of Li-ion batteries[22]. For sodiation of hard carbon, analysis of the scattering signal provides insight into the sodiation mechanism and by combining this with imaging, it can reveal spatial inhomogeneities in the sodiation process across the anode on the micron scale. Spatial inhomogeneities in sodiation across electrodes could lead to parts of the anode being underutilized, affecting the overall capacity and efficiency of the battery. Inhomogeneities in the degree of sodiation over the anode may also lead to localized failure, such as plating of metallic sodium or formation of cracks from localized stress and strain leading to mechanical degradation and loss of electrical conductivity. Furthermore, monitoring macroscopical inhomogeneities can provide information on the sodiation mechanism e.g. by revealing kinetic limitations for the ionic and electronic transport in the electrode by gradients in the sodiation over the anode. In this work we use an approach where we with SWAXS imaging map a full hard carbon electrode in a half-cell in both 2D and 3D to spatially and temporally monitor structural changes during sodiation.

To enable SWAXS imaging of an electrode during cycling, we designed an electrochemical cell with a coated hard carbon electrode in a quartz capillary, Figure. 1a, Figure. S1a. In scanning SWAXS, images are collected by raster scanning the sample over a focused X-ray beam, building up a 2D grid where in each position the scattering pattern of the sample is recorded, Figure 1b, c. From the scattering signal, images can be extracted with the specific contrast originating from a particular scattering feature mapped in each pixel, e.g., the scattering intensity in a chosen q-region. With the high flux offered by synchrotron sources, each scattering pattern is collected within 0.1s, enabling time resolved measurements and a 2D mapping of the full electrode (1.5*0.3 $mm^2$) with a pixel resolution of 15*25 $\mu m^2$ within a few minutes.

In SAXS tomography, multiple 2D images are collected at different projection angles between 0° and 180° to generate a 3D mapping of the electrode, Figure 1d. From the projection images, the scattering of the full electrode volume is reconstructed, generating volumes where each voxel contains the reconstructed scattering signal in that position. This enables both an exact evaluation of the scattering from a specific part of the electrode and from a localized volume of material, with a volume given by the size of the X-ray beam. In this way, SAXS tomography can distinguish between scattering coming from the electrode and from the rest of the cell, enabling a more accurate evaluation of the scattering compared to 2D measurements, with the downside of longer measurement time, on the time scale of a few hours for a tomogram compared to 10 minutes for a similar electrode volume with the same pixel resolution.

In this study we demonstrate the use of a combination of *operando* 2D scanning SWAXS and *ex situ* 3D SAXS tomography to spatially resolve structural changes in model hard carbon electrodes during sodiation and correlate this to the suggested mechanisms of sodiation. *Operando* 2D scanning SWAXS was performed to image the electrode in real time and follow the progression of micropore filling, intercalation and plating over the electrode. Correlating the structural changes with the electrochemistry of the cell suggests that micropore filling is dominating in the plateau region of the voltage profile, while intercalation between stacked graphene layers contribute throughout the full sodiation process. Sodium metal plating is identified in the later stages of the electrochemical process and appears to start growing from the interface of the electrode and the current collector. *Ex situ* 3D SAXS tomography was performed to obtain volume resolved scattering data. With tomography, two *ex situ* cells cycled to different degrees of sodiation were imaged, revealing a higher level of spatial inhomogeneity in the degree of microporosity filling over the electrode at higher degree of sodiation. This demonstrates how SWAXS imaging can provide new insight in the sodiation process of hard carbon anodes for sodium-ion batteries.

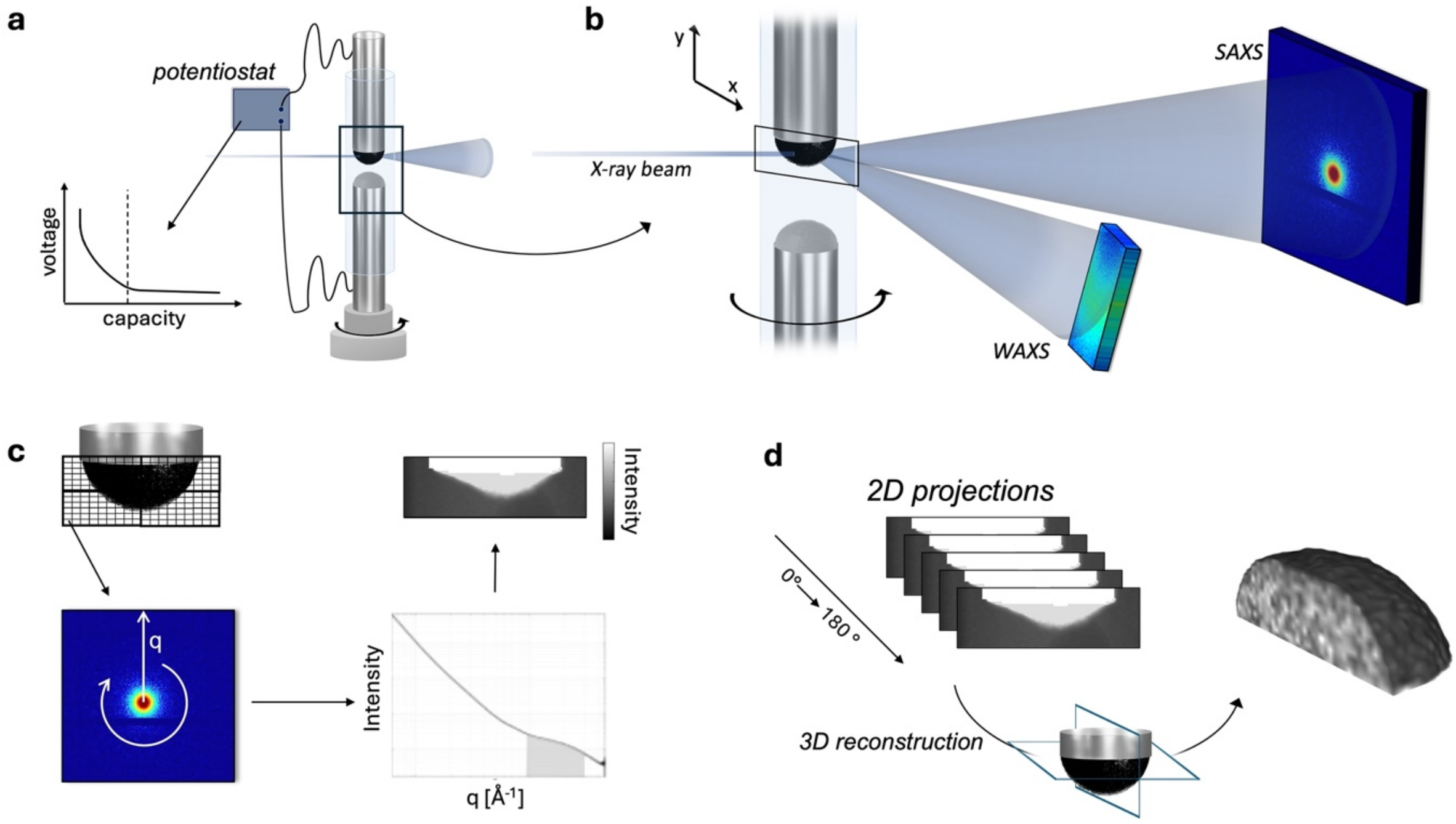

**Figure 1. Schematic of SWAXS imaging of a hard carbon electrode a)** An electrochemical sodium-ion cell is built in a quartz capillary. The cell is positioned on translational and rotational stages, illuminated with a focused X-ray beam and connected to a potentiostat to enable *operando* experiments. **b)** For each illuminated position, a 2D detector image is recorded on a SAXS detector placed 2 m downstream of the sample and a WAXS detector placed below the sample at a distance of 0.6 m, capturing the q-range corresponding to the characteristic scattering properties of hard carbon. **c)** To create a 2D image of the electrode, the sample is raster scanned through the beam, creating a projection of the electrode where each pixel contains the information of the scattering properties in that position. The detector images are azimuthally integrated to retrieve the 1D scattering curves in each pixel. Images of the electrode can then be created by using the scattering intensity in different q-regions as the grey scale of the image. **d)** To perform 3D imaging, the same data acquisition as for the 2D images are applied but from multiple rotation angles to build up a stack of projection images around the sample. These projections can then be used to reconstruct back a 3D volume of the scattering intensity in different q-regions and ultimately produce a 4D volume where the scattering in each voxel position is resolved. To visualize the spatial analysis, the scattering properties in a specific q-region can be selected to create a gray scale contrast in the volume.

# Result and discussion

**2. Structural characterization of the hard carbon anode**
To correlate the sodiation mechanism to the electrochemical profile an analysis model based on the change in scattering intensity during sodiation was applied. Figure 2 shows an overview of the structural differences expected during sodiation of a hard carbon anode and how it is reflected in the scattering intensities. The scattering intensity of hard carbon, $I_{total}$, can be divided into three characteristic q-regions according to equation 1.

$$I_{total} = I_{particles} + I_{micropores} + I_{WAXS} + I_{bkg} \quad \text{eq. 1.}$$

The low-q region, $I_{particles}$ (q < 0.1 Å$^{-1}$), describes the surfaces of hard carbon particles. The mid-q region, $I_{micropores}$ (q ~ 0.1 - 0.5 Å$^{-1}$), is related to the microporosity. The high-q region , $I_{WAXS}$, describes the atomic arrangement of graphene interlayers and inserted sodium in sodiated anodes. Together with the background contribution from other scattering species in the anode, $I_{bkg}$, e.g. inactive binder material, these contributions describe the total scattering of the anode structure. Due to the large difference in scattering intensity between the q-regions (orders of magnitude) the scattering intensity in each region can efficiently be analysed separately.

The scattering intensity in the low-q region, $I_{Particles}$, is described by a power law decay where the slope, $\alpha$, is related to the surface roughness of the particles, eq. 2,

$$I_{particles} = A_{scale}\, q^{-\alpha} \quad \text{eq. 2.}$$

where $A_{scale}$ scales with the amount of material and the exponent, $\alpha$, has a value close to 4 dependent on the surface rougness where $\alpha$ is equal to 4 for a perfectly smooth interface. During sodiation a solid electrolyte interphase (SEI) is formed on the surface of particles from decomposition products of the electrolyte. If material is deposited on the surface, the scattering intensity can be expected to increase as visualised in Figure 2, (magenta panel), and may also lead to a change in the slope if the surface rougness is altered.

The scattering intensity in the mid-q region, $I_{micropores}$, is described by a shoulder which arise from the difference in electron density ($\rho$) between micropores and the carbon matrix. The shape and position of the shoulder will depend on the microstructure of the porosity and is described by a form factor, $P(q)$, and structure factor, $S(q)$, determined by the shape and distribution of the pores in the system, according to equation 3. The intensity is also scaled with the amount of porosity in the material, $K$. If empty pores are filled with sodium, it can be assumed to result in a decreased electron density difference between pores and matrix and consequently a decrease in scattering intensity in the microporous regime[15, 16, 20] as visualised in Figure 2, (green panel).

$$I_{micropores} = K\, (\Delta\rho)^2\, P(q)\, S(q) \quad \text{eq. 3.}$$

The scattering intensity in the high-q region, $I_{WAXS}$, describes the atomic arrangement in the anode. A broad amorphous peak from the interlayer atomic correlation of graphene layers is found around q ~1.6 - 1.7 Å$^{-1}$, $G_{002}$, where the peak position is inversely related to the average interlayer spacing through Bragg's law[20, 23]. The correlation peak is generally referred to as the 002-peak of hard carbon, due to its analogy to the crystalline diffraction peak from stacked graphene layers in graphite[3, 23], although it should be emphasized that hard carbon has a disordered structure. If intercalation occurs, a shift of the peak position towards lower q can be expected due to an increase in average interlayer spacing. Similarly, a decrease in peak intensity can be correlated to the presence of scattering species between the carbon layers which scatter X-rays out of phase with the layers causing destructive interference[23], visualized in Figure 2 (blue panel).

When extensive sodiation occur the atomic correlation of sodium clusters and plating of metallic sodium will be visible at high-q. Plating of metallic sodium will result in a sharp diffraction peak at 2.07 $Å^{-1}$ [20], Figure 2 (yellow panel). Extensive filling of the micropores may also result in amorphous clusters of sodium which is confined within the pores. This results in a broad correlation peak around q ~ 2.1 $Å^{-1}$, $G_{Na_{pseudo-metall}}$, and has been described as pseudo-metallic sodium [20], Figure 2 (orange panel). Hence, the scattering intensity at high q can be described by these three gaussian contributions together with background scattering, $C_{bkg}$, according to equation 4,

$$I_{WAXS} = G_{002}(q_{002}, w_{002}) + G_{Na_{metall}}(q_{Na_{metall}}, w_{Na_{metall}}) + G_{Na_{pseudo-metall}}\left(q_{Na_{pseudo-metall}}, w_{Na_{pseudo-metall}}\right) + C_{bkg} \quad \text{eq. 4.}$$

where the atomic correlation peaks are described by Gaussian peak profiles with their respective peak positions, $(q_{002}, q_{Na_{metall}}, q_{Na_{pseudo-metall}})$ and full width half maximums, $(w_{002}, w_{Na_{metall}}, w_{Na_{pseudo-metall}})$.

The process of adsorption on defects and surfaces are difficult to directly link to a specific change in the scattering intensity of the hard carbon anode as sodium is assumed to be adsorbed both at defects on edges of graphene sheets and on surfaces of the graphene layers within micropores. It can therefore be expected that these processes may induce a slight decrease in intensity of both the microporous region as well as of the 002-peak in the carbon structure from destructive interference from adsorbed sodium on the surfaces of the graphene layers[23, 24].

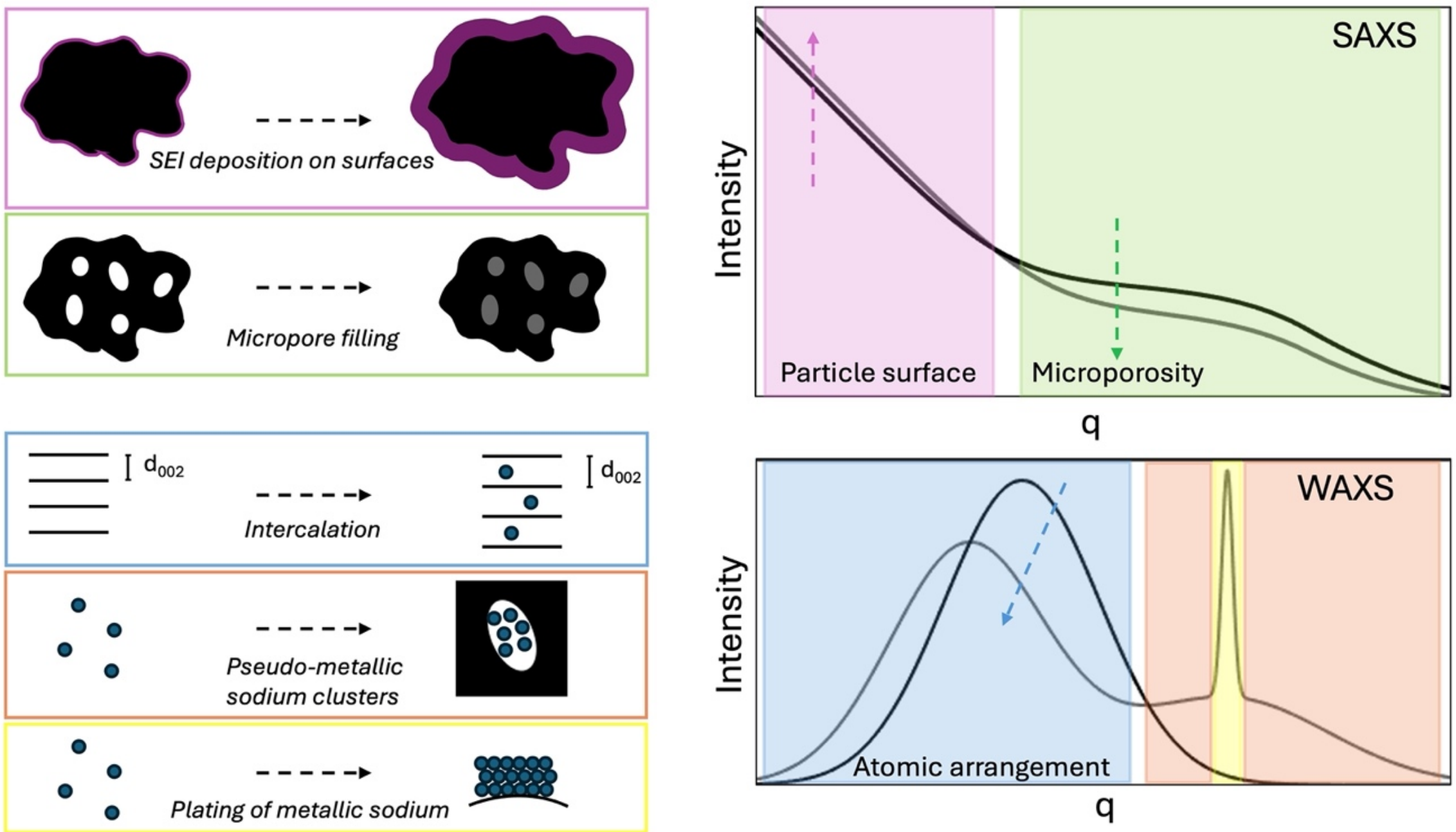

**Figure 2, Schematic of structural changes in the hard carbon anode by sodiation and how it is reflected in the scattering curve.** During sodiation, an SEI layer will be formed on the particle surfaces which may affect the slope and intensity at low-q (magenta). If sodium fills the micropores in the structure, the electron density difference between pores and carbon matrix will decrease and lead to decreased scattering intensity in the microporous region at mid-q (green). Intercalation will lead to a changed shape and intensity of the 002-peak (blue) where a shift towards lower q is expected from an increased interlayer spacing and a decrease in intensity from destructive interference of sodium scattering out of phase with the stacked graphene layers. Plating of metallic sodium will lead to a sharp diffraction peak at high-q (yellow) whereas pseudo-metallic sodium result in a broad amorphous contribution (orange).

For performing a spatial SAXS analysis of the sodiation of hard carbon electrodes a SAXS tomography measurement was first performed on a pristine hard carbon electrode. The electrode was placed in the electrochemical cell to validate the method and to evaluate the homogeneity of a non-sodiated electrode. Figures 3a shows SAXS curves extracted from a few positions marked in Figure 3b to exemplify the scattering signal at different positions. Figure 3b shows two horizontal 2D slices extracted from the tomogram volume with a grayscale of the scattering intensity at $q = 0.1 - 0.5$ Å$^{-1}$ corresponding to scattering from the microporosity in the hard carbon and at $q = 0.075$ Å$^{-1}$ where small inhomogeneities are found across the anode. This scattering contribution is attributed to the binder in the electrode, as supported by previous work showing that the decaying slope from the binder gives an additional scattering contribution at low q [25]. The scattering intensity in the microporous region shows a high homogeneity over the anode, thus the distribution of hard carbon over the electrode is highly homogenous. Figure 3c visualizes a 3D rendering of the electrode derived from the scattering intensity in the microporous region, illustrating the spherical cap shape of the dip coated electrode.

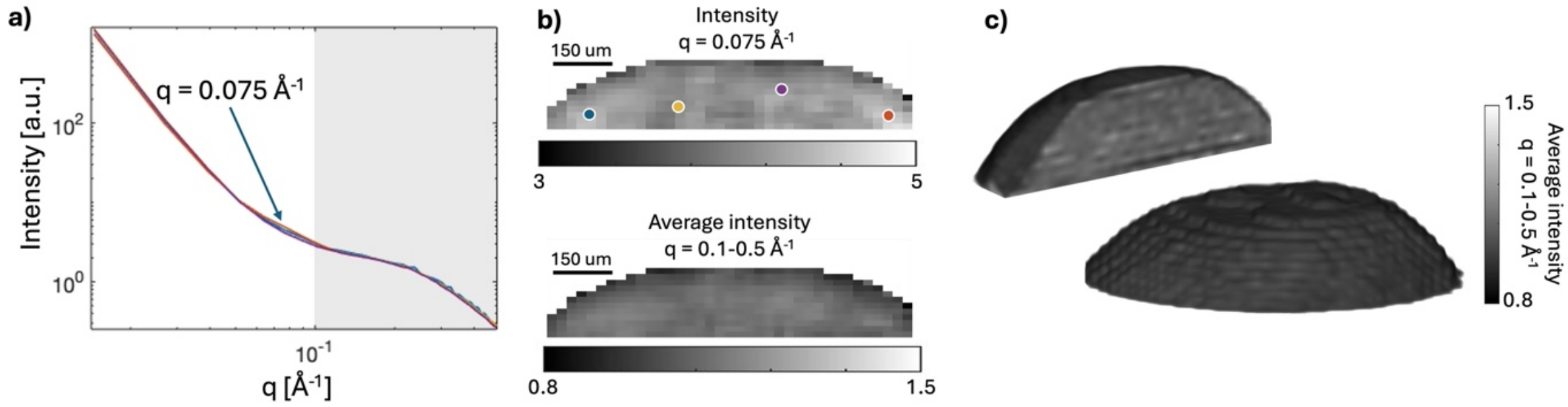

**Figure 3**. **Spatially resolved scattering of a pristine hard carbon electrode**. **a)** SAXS curves reconstructed for the spatial positions marked in **b**. **b)** 2D slices extracted from the tomogram with the scattering intensity at q = 0.075 Å$^{-1}$ and q = 0.1 - 0.5 Å$^{-1}$ **c)** 3D visualization of a hard carbon electrode reconstructed by the scattering intensity in the q-region corresponding to the microporous region. The images display the full electrode and a cross-sectional view of the interior.

### *3. Operando* 2D SWAXS imaging of the sodiation process

Figures 4 -8 show results from the *operando* 2D scanning SWAXS experiment of a hard carbon electrode during galvanostatic sodiation. The applied charge rate (C-rate) was estimated to 0.8C which corresponds to a current density of 0.23 A/g. This estimation was made based on the measured size of the electrode by the SAXS images, as the mass was too small for the accuracy of the scale when coated on the pin. With the estimated volume, the mass of active anode material could be calculated based on the composition, as described in detail in the supporting information, Table S1. The voltage profile, Figure 4a, shows a rapid voltage decrease in the sloping region followed by a plateau region and after about 50 min a decrease in voltage is observed indicating the onset of sodium plating. The cell is a two-electrode cell, and its geometry creates a high internal resistance which contributes to a higher over voltage. This causes a shift of the voltage recorded and the plateau region to appear to have a negative voltage. *Operando* scanning SWAXS images were recorded every 10 min, marked in grey in the voltage profile. Figure 4b shows the averaged scattering intensity over the q-region q = 0.06 - 0.5 Å$^{-1}$, corresponding to the microporous region. Over the first hour of sodiation, the morphology of the electrode is unchanged, as seen by the homogeneous intensity distribution, but in the later stages the intensity distribution becomes progressively more inhomogeneous and the shape of the electrode changes. This coincides with the onset of sodium plating from op 5 see further below, after which the electrode swells and deforms (op 7) and finally starts to delaminate (post).

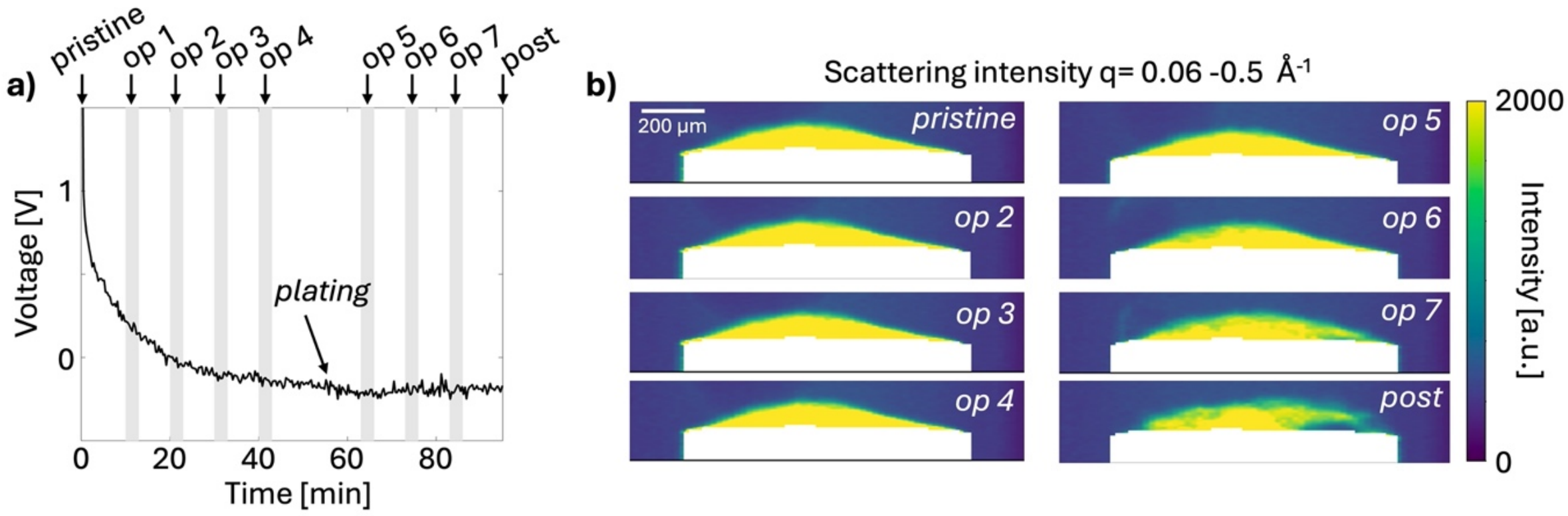

**Figure 4.** ***Operando* 2D SWAXS imaging of the electrode a)** Galvanostatic voltage profile of the sodiation of the hard carbon electrode. The cell is cycled with a current of 14 µA corresponding to an estimated C-rate of 0.8C. Grey lines indicate points of collection of scanning SWAXS images during the *operando* experiment. Each operando scan is described by the consequent scan number, op #. **b)** Scanning SWAXS images of the averaged intensity in the q-region (q = 0.06 - 0.5 $Å^{-1}$) where hard carbon scatters strongly and the electrode appears bright. The electrolyte surrounding the electrode does not scatter in this q-region and appears blue. The current collector is masked in white.

Figure 5a shows the SAXS curves where the intensity has been averaged over the full projected area of the electrode for the first five *operando* scans. The SAXS curves display the characteristic scattering features of hard carbon and an additional contribution around q = 0.075 $Å^{-1}$. This contribution is attributed to the inactive materials in the electrode, including the binder, and remains stable during cycling and does not affect the analysis of structural changes in the hard carbon during sodiation evaluated in the experiment. Two main changes are observed in the SAXS signal during sodiation. A decrease in intensity in the microporous region (q = 0.12 - 0.4 $Å^{-1}$), and an increase in intensity in the particle surface region (q = 0.0045 - 0.007 $Å^{-1}$), Figure 5a grey highlighted regions. The time evolution of the average intensities across the anode in these two q-regions is shown in Figures 5b. Initially, the intensity in the microporous region increases slightly followed by a pronounced decrease in intensity. This change in intensity can be attributed to the changes in electron density difference between the micropores and the carbon matrix, eq. 3. The initial increase can be related to an increased electron density of the carbon matrix due to sodium intercalation in the carbon structure[15] which implies that sodium first start to intercalate. As the sodiation progresses further, the intensity decreases, indicating a decreased electron density difference between pores and carbon matrix, suggesting filling of micropores with sodium. When the electrochemical profile reaches the plateau region, the intensity decreases more rapidly reflecting that pore filling is now the dominating mechanism, in agreement with previous reports in literature [8, 13-17]. The results reveal an intensity decrease of ~35%, (proportional to the electron density difference between the pores and the carbon matrix), which is comparable to previous results in literature (e.g. ~25% decrease after complete sodiation in the work of Iglesias et al. [15])

In the low-q region related to scattering from the hard carbon particles surfaces, the intensity increases continuously during sodiation however more rapidly initially, Figure 7b. This increase is less straightforward to directly correlate to a sodiation mechanisms and is more likely related to SEI formation on the surface of the particles. The more rapid initial intensity increase is consistent with SEI formation which is formed at higher voltages in the initial stages of sodiation[14].

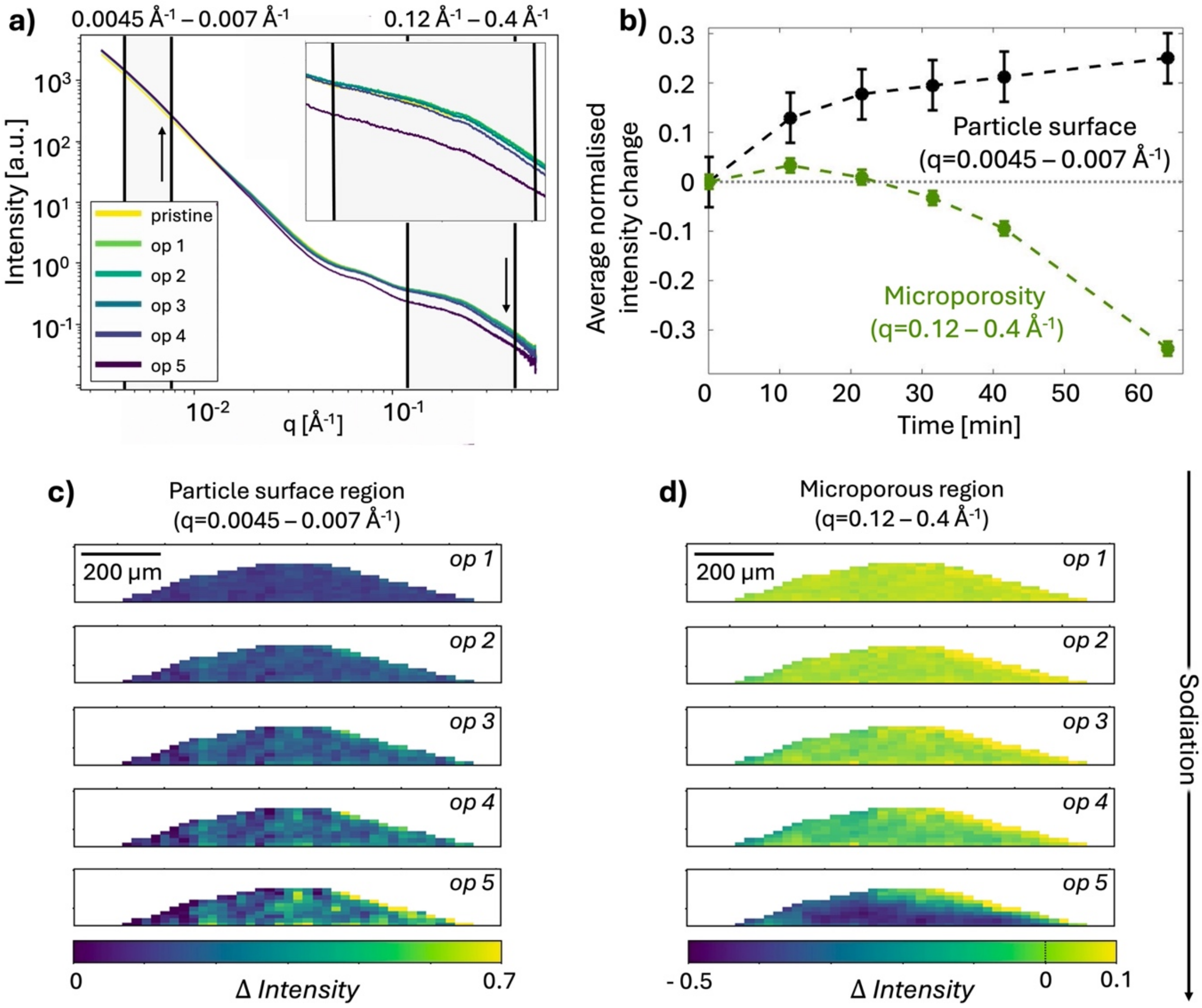

**Figure 5. Structural changes in the SAXS region**. a) SAXS curves of the averaged intensity over the full projected area of the electrode for the first five operando scans. The q-ranges used to visualize the scattering intensity in the particle surface region and the microporous region are marked by black lines. Inset shows a magnified view of the q-range corresponding to scattering from micropores. **b)** Plot of the changes in intensity in the selected q-ranges (marked by black lines in a) during the electrochemical process averaged over the full electrode. Error bars correspond to the standard deviation of the intensity across the anode within the first difference map. **c)** Difference maps of the averaged intensity in the particle surface region (0.0045- 0.007 $Å^{-1}$) and **d)** microporous region (0.12 - 0.4 $Å^{-1}$). Difference maps are created by subtracting the intensity of the pristine scan from the *operando* scan followed by division by the intensity of the pristine electrode, to normalize the signal from thickness variations over the electrode and highlight changes in intensity in each pixel. The area surrounding the electrode is masked in white.

Figures 5c,d show scanning SAXS images of the scattering intensity in the microporous (q = 0.12 - 0.4 $Å^{-1}$) and particle surface (q = 0.0045 - 0.007 $Å^{-1}$) regions, respectively. The 2D images are displayed as difference maps between the intensity in the selected q- regions of the *operando* image (op #) and in the pristine image, according to equation 5. The difference maps are created by subtracting the average intensity in the selected q-region of the pristine electrode from the intensity in the *operando scans* in each pixel and normalizing by the intensity from the pristine electrode, to emphasize the differences during sodiation and correct for thickness variations in the 2D projections. Figure S2 shows how the scattering curves change at a few selected points in the sample.

$$Difference\ map = \frac{\bar{I}_{op\,\#}^{selected\ q-region} - \bar{I}_{pristine}^{selected\ q-region}}{\bar{I}_{pristine}^{selected\ q-region}} \qquad \text{eq. 5.}$$

During sodiation, the intensity in the microporous region first increases rather homogenously over the electrode (op 1 and 2 in Figure 5d). As the sodiation process continues (op 3-5), inhomogeneities arise and a decreased intensity can be observed over almost the entire electrode. In particular, the region close to the current collector shows a clear intensity decrease, indicating a higher degree of pore filling occurring in this part of the electrode. A larger degree of sodiation close to the current collector indicates that the electrode does not experience ion-depletion, which would be expected to result in the opposite gradient with a higher degree of sodiation close to the surface of the electrode and the electrolyte. The radial gradient with higher degree of sodiation closer to the current collector can instead be a result of a low electronic conductivity in the electrode leading to faster kinetics closer to the current collector. In the particle surface region, an increased inhomogeneity in the intensity difference is also observed throughout the sodiation process. The spatial changes in this region appear not to be correlated with the spatial changes in the microporous region, suggesting that the origins of the structural changes are not correlated.

Figure 6a shows the WAXS curves where the intensity has been averaged over the full projected area of the electrode for the first five *operando* scans. In the first scans the WAXS curve contains the scattering signal from the broad 002-peak from hard carbon as well as a peak at q = 1.45 $Å^{-1}$ attributed to the binder, Figure S3. In the fifth operando scan (op 5), a broad contribution at q = 2.0 - 2.3 $Å^{-1}$ appears which can be related to the formation of pseudo-metallic sodium[20], Figure 6b inset, resulting from extensive micropore filling leading to formation of confined sodium clusters in the micropores[20, 26]. A small trace of a sharp diffraction peak of metallic sodium can also be identified at 2.08 $Å^{-1}$, showing the onset for plating in the anode.

To extract the peak position and define the contribution from pseudo-metallic sodium the WAXS curves were modelled according to equation 4, described in more detail in the supporting information. Figure S4, Table S2. During sodiation the 002-peak decreases in intensity and the peak position shifts to lower q-values, which both are indicators for sodium intercalation[20]. The peak shift of the 002-peak was converted to the average interlayer spacing for each operando scan with Braggs law ($d_{002}=2\pi/q_{002}$). The results are shown in Figure 6b, showing a continuous increase in average interlayer spacing throughout sodiation. The change in interlayer spacing correlates well with the decrease in peak intensity, Figure 6b, suggesting that intercalation occurs throughout the sodiation process. This agrees with the SAXS results, which showed an increased electron density of the carbon matrix related to intercalation early in the sodiation process, before micropore filling starts dominating the SAXS signal.

Figure 6c shows the difference map of the averaged intensity of the 002-peak over the electrode. The signal to noise ratio is rather low in each pixel in this q-range due to the strong background from the electrolyte, but a slightly larger decrease in intensity can be observed in the outer left part of the electrode as well as closer to the current collector. This shows that also intercalation occurs favorably closer to the current collector, as observed for the microporous region in SAXS that identified that a larger fraction of micropores is filled at this part of the anode which could indicate low electronic conductivity in the electrode.

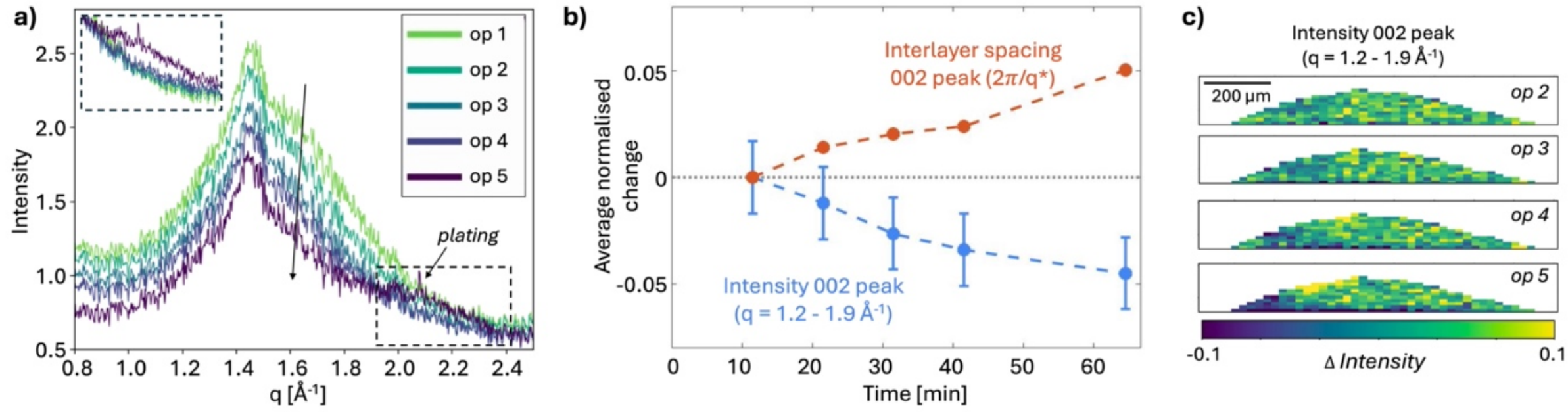

**Figure 6**. **Structural changes in the WAXS region a)** WAXS curves of the averaged intensity over the full projected area of the electrode for the first five operando scans. Inset shows the curves normalized for the change in intensity of the 002-peak revealing a contribution from pseudo-metallic sodium in the range q= 2.0 - 2.3 $Å^{-1}$ which appear in the 5th operando scan. **b)** Plot of the changes in intensity in the q-range of the 002-peak (q = 1.2 – 1.9 $Å^{-1}$) and the average interlayer spacing of graphene layers derived from the 002-peak position. Error bars correspond to the standard deviation of the intensity variation in the anode within the first difference map for the intensity of the 002-peak. **c)** Difference map of averaged intensity of the 002-peak (q = 1.2 – 1.9 $Å^{-1}$). The area surrounding the electrode is masked in white.

Figure 7a shows the difference map of the intensity of the q-region corresponding to the broad peak from pseudo-metallic sodium. To exclude contribution from plated sodium, the intensity of the q-region from the contribution from pseudo-metallic was selected as q = 2.1 - 2.3 $Å^{-1}$. Similar to the SAXS results for micropore filling, the highest signal for pseudo-metallic sodium is here found close to the current collector, indicating extensive pore filling in this region of the electrode. Figure 7b shows a scatter plot of the intensity difference in each pixel of the electrode within the WAXS region for pseudo-metallic sodium (y-axis) and the SAXS region for microporosity filling (x-axis). The plot visualizes how the higher contribution of the broad pseudo-metallic peak appears where there is also a low intensity in the SAXS microporous region (top left area, Figure 7b) which shows that the micropores indeed are extensively filled at the onset of formation of pseudo-metallic sodium. However, the presence of micropore filling and pseudo-metallic sodium are not fully correlated in the images and rather suggests a relationship where a high micropore filling is needed for pseudo-metallic sodium to form, but regions with high micropore filling does not directly imply formation of pseudo-metallic sodium.

To summaries the interpretation of the structural changes related to the different sodiation mechanisms Figure 7c-d shows an overview of the changes observed in the electrode during sodiation. Figure 7d shows a combined plot of the normalized changes in intensity in the different q-regions and Figure 7c the electrochemical profile of the cell. The results visualize how the intensity and derived interlayer spacing of the 002-peak, related to sodium intercalation, changes continuously during the electrochemical process whereas the scattering related to micropores in the carbon matrix decreases faster in the plateau region in the later stages of sodiation. This indicates that intercalation occurs continuously throughout the sodiation process while filling of microporosities dominates in the later stages of the electrochemical process which is also supported by the appearance of a peak for pseudo-metallic sodium in the fifth operando scan, Figure 6 and S4.

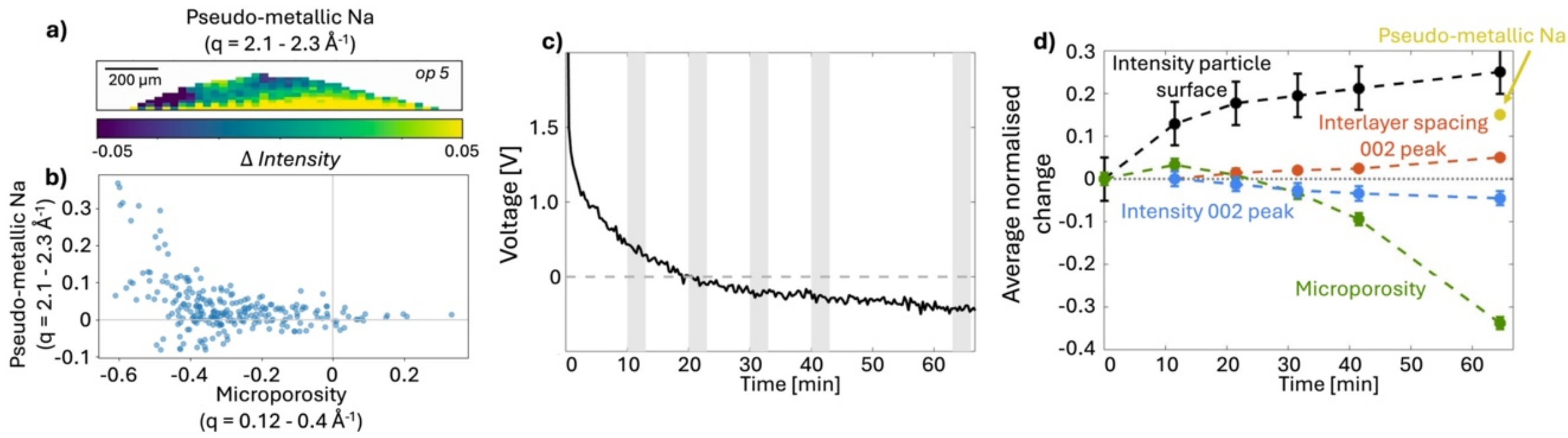

**Figure 7**. **Overview of the changes in scattering signal over the electrode during sodiation a)** Difference map of the average intensity in the q-region of pseudo-metallic sodium (q = 2.1 - 2.3 Å$^{-1}$). The area surrounding the electrode is masked in white. **b)** Comparison of the intensity change in the q-regions related to microporosity filling from SAXS (x-axis, q = 0.12 - 0.4 Å$^{-1}$) and WAXS (y-axis, (q = 2.1 - 2.3 Å$^{-1}$). **c)** Voltage profile of the sodiation process of the first 60 min before excessive sodium plating **d)** Overlay of the line plots of the average intensity in selected q-regions corresponding to the different mechanisms of sodiation during the electrochemical process. The plot visualizes the changes in the particle surface region, q = 0.0045 - 0.007 Å$^{-1}$ (black), microporous region q = 0.12 - 0.4 Å$^{-1}$ (green) and the 002-peak q = 1.2 - 1.9 Å$^{-1}$ (blue) as well as the change in average interlayer spacing of graphene layers (orange). The appearance of pseudo-metallic sodium, q = 2.1 - 2.3 Å$^{-1}$, after 60 min is illustrated by the yellow marker. Error bars correspond to the standard deviation of the intensity across the anode within the first difference map.

After 65 min (op 5), signatures of Na-plating can be detected in the WAXS curve by the appearance of a diffraction peak at 2.08 Å$^{-1}$ from metallic sodium, Figure 8a. Figure 8b shows the spatial distribution of the intensity corresponding to the diffraction peak over the electrode, where a yellow color indicates a presence of metallic sodium. The plating starts close to the current collector and progressively grows from the bottom of the electrode. The excessive plating in the last scans can be correlated to the swelling, deformation and delamination of the electrode from the current collector observed in Figure 4b. These results demonstrate a clear preference of plating to occur first close to the current collector. One can note that also a high degree of micropore filling and, in particular, formation of pseudo-metallic sodium occurs in the same region close to the current collector. This may suggest that the highly sodiated sites with pseudo-metallic sodium can act as precursors to sodium plating.

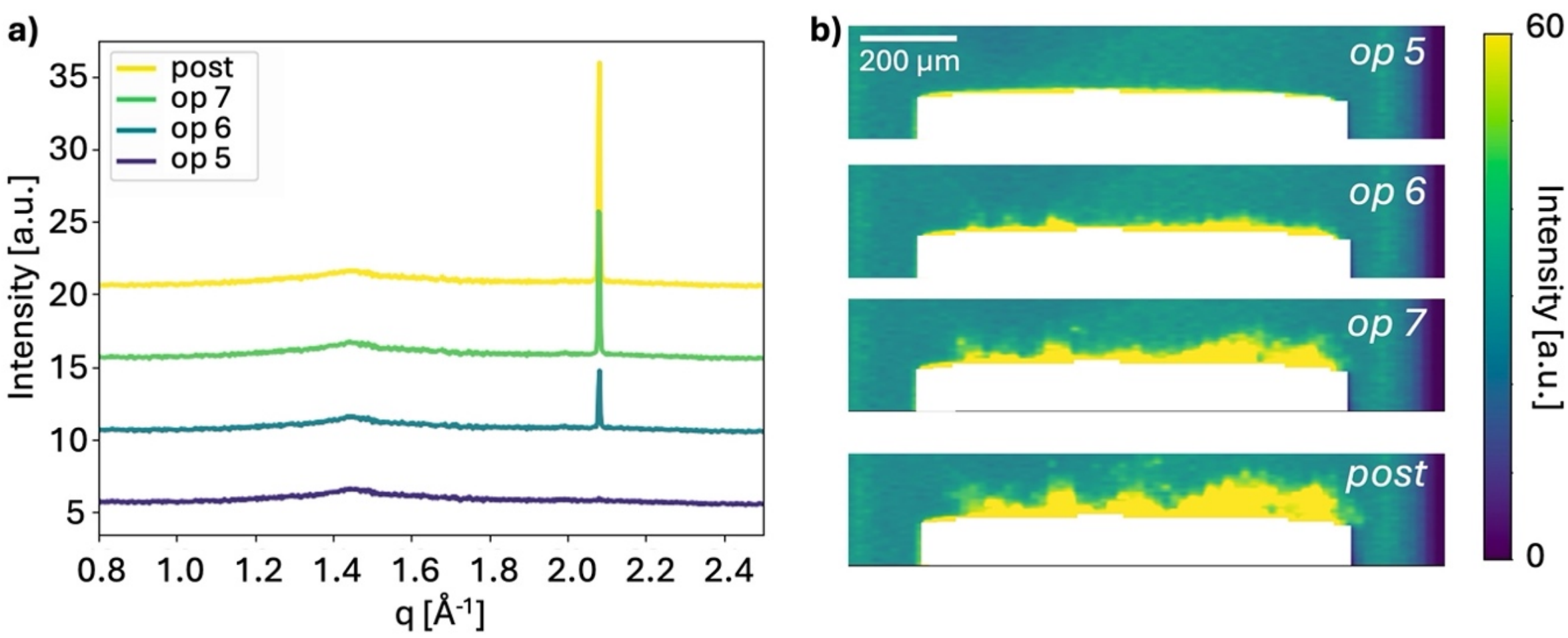

**Figure 8. Spatial and temporal progression of sodium plating in the electrode a)** Average WAXS curves of the electrode in the later stages of sodiation of the electrode (op 5-7, post), offset for visibility. **b)** Scanning WAXS images of the intensity distribution of the peak for metallic sodium (2.08 Å$^{-1}$), sodium plating corresponds to a high intensity. The current collector is masked in white.

**4. 3D SAXS tomography of sodiated electrodes**

SAXS tomography was performed on two sodiation cells to analyze heterogeneities across the entire anode volume on the micron scale. Due to the long acquisition time for a tomogram, 3D SAXS analysis was not compatible with *operando* measurements and were instead performed on *ex situ* samples. The *ex situ* samples were extracted from two cells cycled to different degree of sodiation. For the first electrode, sodiation was stopped in the sloping region (100 min) after cycling with an estimated C-rate of 0.3C whereas for the second electrode sodiation was stopped far in the plateau region (200 min) after cycling with an estimated C-rate of 0.2C, Figure 9a. These two cases will further on be referred to as the electrode of low and high degree of sodiation and be compared to the pristine electrode, Figure 3.

Figure 9 provides an overview of the results from the SAXS tomography. Figures 9b, c shows scattering curves from two selected voxels for each sample to exemplify the scattering from the two different electrodes. The average scattering curve from the pristine electrode is included for comparison (black curves) to emphasize the differences in scattering as a result of sodiation. As expected, the intensity in the microporous region (0.2 - 0.5 $Å^{-1}$), characteristic for pore filling, decreases with an increasing degree of sodiation. A decrease in slope in the low q-region is found for both electrodes, and an increased intensity contribution around 0.075 $Å^{-1}$ is found in the electrode with a high degree of sodiation. The contribution at 0.075 $Å^{-1}$ is here most likely due to SEI formation on the surface of the hard carbon particles as well as residues of salt and electrolyte in the *ex situ* cells which can create particles on the surface of the electrode on that length scale.

To evaluate spatial inhomogeneities in sodiation across the electrode, the microporous q-region was examined in more detail. Figures 9d-f show 2D slices of the averaged intensity in the microporous region (0.2 - 0.5 $Å^{-1}$) extracted horizontal and vertically from the center of the tomograms. As each voxel contains the scattering signal from a defined volume of material, the scattering intensity is directly comparable between voxels and between the different samples. The pristine sample, Figure 9d, therefore serves as a reference for the intensity distribution and homogeneity of a hard carbon electrode. Compared to the pristine sample, the sodiated electrodes show decreased intensity in the microporous region, and increased inhomogeneity in intensity distribution across the electrode. Figure 9g displays histograms of the averaged intensity in the three electrode volumes. The histogram visualizes the decrease in intensity between the electrodes as well as an increased width of the histogram, illustrating the increased inhomogeneity in micropore filling in the highly sodiated electrode. The results show a decrease in intensity in both sodiated electrodes, suggesting that some pore filling already occurs at low degree of sodiation, within the sloping voltage profile. One explanation to the presence of sodium in the micropores in the sample sodiated to a low degree of sodiation could be that adsorption on defects within the pores occur in the early stage of sodiation and may therefore also affect the scattering intensity in this q-range. With increased degree of sodiation, the average intensity decreases further, and the width of the histogram increases.

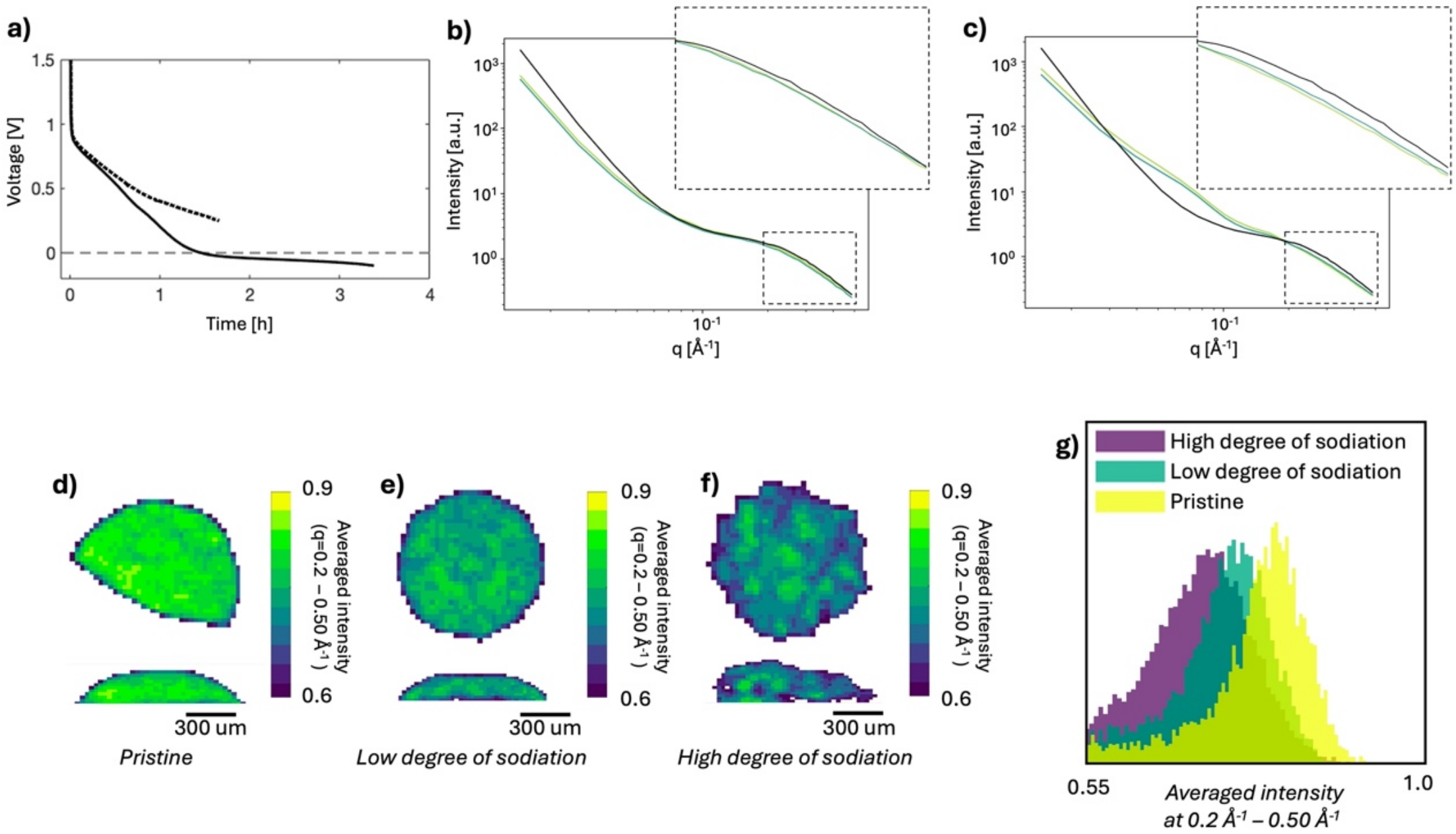

**Figure 9. *ex situ* 3D SAXS tomography of hard carbon electrodes. a)** Voltage profile of the *ex situ* cells with low degree of sodiation (dashed line) and high degree of sodiation (full line) discharged with an estimated C-rate of 0.3C for 100 min and 0.2C for 200 min, respectively. SAXS curves in a few selected voxels in the electrode with b) low degree of sodiation and c) high degree of sodiation. The reference to the scattering intensity of the pristine electrode is shown in black. 2D slices extracted horizontally and vertically from the tomograms of the averaged intensity in the microporous region (0.2 - 0.5 $Å^{-1}$) of the **d)** pristine hard carbon electrode, **e)** the electrode with low degree of sodiation and **f)** the electrode with high degree of sodiation. **g)** Histograms of the averaged intensity of the microporous q-region for the three electrodes. Due to the outer voxels at the electrode surface being only partly filled with electrode, a lower intensity value is observed here which is reflected in the tails of the histogram and is not related to the sodiation process but rather the shape of the electrode.

## 5. Conclusion

In this paper we demonstrate SWAXS imaging as a characterization method to in detail investigate the mechanisms of sodiation of hard carbon electrodes. By imaging the local scattering signal over the hard carbon anode, structural changes are spatially resolved and correlated to the mechanisms of sodiation and the electrochemical process. With scanning SWAXS, we perform *operando* 2D imaging of a hard carbon anode during sodiation, monitoring the structure in real time. To analyze the data, we present a model for performing scattering intensity difference mapping across the electrode and relate the structural differences to the sodiation process. This reveals that sodiation through micropore filling is the more dominating mechanism in the later stages of sodiation, while intercalation occurs continuously during sodiation. Plating is shown to initiate from the interface of the electrode and current collector where also a higher degree of micropore filling is found. With SAXS tomography, we image the scattering in the anode in 3D, revealing an increased level of inhomogeneity at a higher degree of sodiation with regions of different degrees of micropore filling.

The application of SWAXS imaging presents opportunities for detailed studies of the sodiation in hard carbon anodes, not only by spatially resolved analysis, but also as a method for decoupling the contributions of different components in a cell, enabling more accurate scattering analysis in *in situ* environments. By using a multiscale imaging approach, SWAXS imaging enables mapping of structural changes during charging to also reveal heterogeneities on the micron scale, over a macroscopic electrode, which can give insights into the functionality, performance and failure mechanism of hard carbon electrodes. Thus, our approach paves way for a deeper understanding of the storage mechanism in hard carbon anodes.

# Methods

**Electrochemical cell preparation**
A dedicated electrochemical cell for *operando* SWAXS imaging was built using a quartz capillary with an internal diameter of 1.5 mm and wall thickness of 150 µm. Quartz capillaries were used to minimize the SAXS scattering background. Stainless steel pins were used as current collectors, polished to obtain a flat substrate, and the hard carbon electrode was prepared by dip coating the flat tip of the stainless-steel pin in a hard carbon slurry and dried at 80 °C in vacuum overnight. The slurry was made of 85 wt.% hard carbon (Kureha Battery Materials Japan), 10 wt.% polyvinylidene fluoride (PVDF) binder (Sigma Aldrich) and 5 wt.-% carbon additives (Carbon black Super P®, ThermoFisher), dispersed in N-methyl-2-pyrrolidone (NMP), (Sigma Aldrich). The cell was assembled by inserting the pin with the coated hard carbon electrode, filling it with electrolyte, 1M $NaPF_6$ in EC: PC (Ethylene carbonate: Propylene carbonate) 1:1 by weight (E-Lyte Innovation), and inserting a pin with Na metal (dry stick, Thermo scientific chemicals) attached to it as counter electrode in the glass capillary. The cell was sealed with UV-curable glue. Schematics and photographic images of the cell are shown in Figure S1. For *ex situ* tomography experiments the pin with the hard carbon electrode was first cycled within the capillary cell and then transferred to an empty capillary without electrolyte after cycling to remove excess scattering and absorption of electrolyte, as well as prevent evolution of the sample after cycling. The dip coating resulted in a variation in mass between electrodes below the accuracy of the scale and the mass of active material (hard carbon) was instead calculated based on the volume determined from SWAXS imaging and the composition of the slurry. The calculations are shown in the supporting information. From the mass of active material an applied C-rate of approximately 0.8C was estimated for the operando cell, while for the *ex situ* cells a C-rate of 0.3C and 0.2C was estimated for the electrodes cycled to a low and high degree of sodiation, respectively (Table S1).

**SWAXS and SWAXS imaging measurement**
Scanning SWAXS and SAXS tomography of the electrodes were performed at the cSAXS beamline (X12SA) at the Swiss Light Source, Paul Scherrer Institute (Switzerland). The *operando* experiment and *ex situ* tomography experiments were performed in two different experiments but using the similar setup. An X-ray beam (13.2 keV photon energy) was focused to a spot size of approximately 25 x 15 µm$^2$. An evacuated flight tube was placed between the sample and the detectors to reduce air scattering and absorption. Scattering patterns were simultaneously measured using a Eiger 9M detector for SAXS in the *operando* experiment and a Pilatus 2M detector in the *ex situ* tomography experiment. A Pilatus 300kw detector oriented as a vertical strip were placed below the sample for WAXS. The sample to detector distance was approximately 2 m for SAXS and 0.6 m for WAXS. The q-ranges covered by the two detectors were determined with calibration measurements of silver behenate (SAXS) and lanthanum hexaboride (WAXS).

The samples were mounted on rotational and translational stages with the hard carbon electrode on the top to have free path for the scattered X-rays to reach the WAXS detector placed underneath, schematic Figure 1 and photographic images in Figure S1. Each scanning SWAXS image and the projections in the tomograms were scanned with a step size of 25 and 15 µm in the horizontal and vertical directions, respectively, using an exposure time of 0.1 s. For the tomograms 72 projections were measured. To check for radiation damage, due to the high intensity X-ray exposure of multiple projections being measured in the SWAXS imaging experiments, the first projection at 0° was remeasured after a full tomogram and compared to the first projection. No differences were detected in the scattering signal after the X-ray exposure, Figure S5.

**Reconstruction of tomograms and data analysis**

The two-dimensional scattering patterns were azimuthally integrated to retrieve one-dimensional scattering curves with the Matlab *cSAXS scanning SAXS package* developed by the CXS group, Paul Scherrer Institut (Switzerland)[27]. Each measurement was transmission corrected by the diode data and background subtraction was made using the scattering intensity in an area of the cell above the electrode containing only electrolyte. The q-regions were selected to minimise overlap from different contributions. To analyze the 2D images collected during *operando* sodiation the scattering intensity in the selected q-regions were averaged in each pixel. The current collector was masked out of the image based on having a transmission signal lower than 0.1, as the steel pin is highly absorbing. The area surrounding the electrode in the cell was masked based on the scattering intensity in the range q = 0.12 – 0.3 $Å^{-1}$, where the hard carbon anode scatters strongly while there is no scattering from electrolyte. In the masked 2D images a difference map of the change in intensity in the selected q-regions were created by calculating the difference between the *operando* image (op #) collected during sodiation and the pristine image. The difference maps are created by subtracting the average intensity in the selected q-region of the pristine electrode from the intensity in the *operando* scans in each pixel and normalizing by the intensity from the pristine electrode to correct for thickness variations inherent in the 2D projections, according to equation 5. The average intensity difference across the anode were plotted as a function of time to relate the structural changes to the sodiation mechanism while the difference maps were used to visualize spatial inhomogeneities on the micron scale across the anode. Error bars for the averaged intensity difference in each q-region was determined by the standard deviation of the intensity in all pixels across the anode in the first scan, serving as a reference for the noise variation in the selected q-region across the anode.  To extract the peak position and define the contribution from pseudo-metallic sodium the WAXS curves were modelled according to equation 4. To derive the interlayer spacing from the peak shift of the 002-peak Bragg's law was applied. The fitting is described in more detail and results are shown in the supporting information.

To perform tomographic reconstructions an iterative reconstruction algorithm developed by Liebi et. al. [28] was applied where the intensity of the scattering is optimized with the use of spherical harmonics as commonly applied for SAXS tensor tomography. The scattering curve was divided in 39 linearly spaced q-bins and the 2D scattering projections were transmission corrected  with the diode data and aligned using the scattering intensity in the range q = 0.12 – 0.3 $Å^{-1}$ [29]. The electrode volume was masked by the scattering intensity in the same q-range. The generated 3D SAXS tomograms were filtered using a mean 3D filter with a kernel of 2*2*2 pixels to smoothen noise in the visualisations. The homogeneity of the electrodes was compared by creating a histogram over the greyscale values of the chosen scattering intensities in the masked region of the electrode volume.

# Acknowledgement

The authors acknowledge the Paul Scherrer Institute, Villigen PSI, Switzerland for provision of synchrotron radiation beamtime at the beamline cSAXS of the SLS. The work was performed in part at Chalmers Analysis Laboratory (CMAL). This work was supported by the Area of Advance Nano at Chalmers University of Technology through an excellence initiative PhD student position. We acknowledge Magnus Karlsteen at Chalmers University of Technology for assisting with 3D printing of sample holders for the experimental setup at cSAXS. LN, LR and ML received funding from the European Research Council Starting Grant MUMOTT (ERC-2020-StG 949301), funded by the European Union. MC and CA have received funding from the European Union's Horizon 2020 research and innovation programme under the Marie Skłodowska-Curie grant agreement No. 884104. Views and opinions expressed are however those of the author(s) only and do not necessarily reflect those of the European Union or the European Research Council Executive Agency. Neither the European Union nor the granting authority can be held responsible for them.

# References

(1) Dou, X.; Hasa, I.; Saurel, D.; Vaalma, C.; Wu, L.; Buchholz, D.; Bresser, D.; Komaba, S.; Passerini, S. Hard carbons for sodium-ion batteries: Structure, analysis, sustainability, and electrochemistry. *Materials Today* **2019**, *23*, 87-104.
(2) Delmas, C. Sodium and sodium-ion batteries: 50 years of research. *Advanced Energy Materials* **2018**, *8* (17), 1703137.
(3) Saurel, D.; Segalini, J.; Jauregui, M.; Pendashteh, A.; Daffos, B.; Simon, P.; Casas-Cabanas, M. A SAXS outlook on disordered carbonaceous materials for electrochemical energy storage. *Energy Storage Materials* **2019**, *21*, 162-173. DOI: 10.1016/j.ensm.2019.05.007.
(4) Wen, Y.; He, K.; Zhu, Y.; Han, F.; Xu, Y.; Matsuda, I.; Ishii, Y.; Cumings, J.; Wang, C. Expanded graphite as superior anode for sodium-ion batteries. *Nature communications* **2014**, *5* (1), 4033.
(5) Rodriguez-Palomo, A.; Lutz-Bueno, V.; Cao, X.; Kádár, R.; Andersson, M.; Liebi, M. In situ visualization of the structural evolution and alignment of lyotropic liquid crystals in confined flow. *Small* **2021**, *17* (7), 2006229.
(6) Xie, F.; Xu, Z.; Guo, Z.; Titirici, M.-M. Hard carbons for sodium-ion batteries and beyond. *Progress in Energy* **2020**, *2* (4), 042002.
(7) Simone, V.; Boulineau, A.; De Geyer, A.; Rouchon, D.; Simonin, L.; Martinet, S. Hard carbon derived from cellulose as anode for sodium ion batteries: Dependence of electrochemical properties on structure. *Journal of energy chemistry* **2016**, *25* (5), 761-768.
(8) Stratford, J. M.; Kleppe, A. K.; Keeble, D. S.; Chater, P. A.; Meysami, S. S.; Wright, C. J.; Barker, J.; Titirici, M.-M.; Allan, P. K.; Grey, C. P. Correlating local structure and sodium storage in hard carbon anodes: insights from pair distribution function analysis and solid-state NMR. *Journal of the American Chemical Society* **2021**, *143* (35), 14274-14286.
(9) Chen, X.; Tian, J.; Li, P.; Fang, Y.; Fang, Y.; Liang, X.; Feng, J.; Dong, J.; Ai, X.; Yang, H. An overall understanding of sodium storage behaviors in hard carbons by an "adsorption-intercalation/filling" hybrid mechanism. *Advanced Energy Materials* **2022**, *12* (24), 2200886.
(10) Alvin, S.; Yoon, D.; Chandra, C.; Cahyadi, H. S.; Park, J.-H.; Chang, W.; Chung, K. Y.; Kim, J. Revealing sodium ion storage mechanism in hard carbon. *Carbon* **2019**, *145*, 67-81.
(11) Xiao, B.; Rojo, T.; Li, X. Hard carbon as sodium-ion battery anodes: progress and challenges. *ChemSusChem* **2019**, *12* (1), 133-144.
(12) Chen, X.; Liu, C.; Fang, Y.; Ai, X.; Zhong, F.; Yang, H.; Cao, Y. Understanding of the sodium storage mechanism in hard carbon anodes. *Carbon Energy* **2022**, *4* (6), 1133-1150.
(13) Kim, H.; Hyun, J. C.; Kim, D. H.; Kwak, J. H.; Lee, J. B.; Moon, J. H.; Choi, J.; Lim, H. D.; Yang, S. J.; Jin, H. M. Revisiting Lithium-and Sodium-Ion Storage in Hard Carbon Anodes. *Advanced Materials* **2023**, *35* (12), 2209128.
(14) Weaving, J. S.; Lim, A.; Millichamp, J.; Neville, T. P.; Ledwoch, D.; Kendrick, E.; McMillan, P. F.; Shearing, P. R.; Howard, C. A.; Brett, D. J. Elucidating the sodiation mechanism in hard carbon by operando raman spectroscopy. *ACS Applied Energy Materials* **2020**, *3* (8), 7474-7484.
(15) Kitsu Iglesias, L.; Antonio, E. N.; Martinez, T. D.; Zhang, L.; Zhuo, Z.; Weigand, S. J.; Guo, J.; Toney, M. F. Revealing the Sodium Storage Mechanisms in Hard Carbon Pores. *Advanced Energy Materials* **2023**, *13* (44), 2302171.
(16) Stevens, D.; Dahn, J. An in situ small-angle X-ray scattering study of sodium insertion into a nanoporous carbon anode material within an operating electrochemical cell. *Journal of The Electrochemical Society* **2000**, *147* (12), 4428.

(17) Zhang, B.; Ghimbeu, C. M.; Laberty, C.; Vix-Guterl, C.; Tarascon, J. m. Correlation between microstructure and Na storage behavior in hard carbon. *Advanced Energy Materials* **2016**, *6* (1), 1501588.
(18) Zhou, H.; Song, Y.; Zhang, B.; Sun, H.; Khurshid, I. A.; Kong, Y.; Chen, L.; Cui, L.; Zhang, D.; Wang, W. Overview of electrochemical competing process of sodium storage and metal plating in hard carbon anode of sodium ion battery. *Energy Storage Materials* **2024**, 103645.
(19) Jian, Z.; Xing, Z.; Bommier, C.; Li, Z.; Ji, X. Hard carbon microspheres: potassium-ion anode versus sodium-ion anode. *Advanced Energy Materials* **2016**, *6* (3), 1501874.
(20) Morikawa, Y.; Nishimura, S. i.; Hashimoto, R. i.; Ohnuma, M.; Yamada, A. Mechanism of sodium storage in hard carbon: an X-ray scattering analysis. *Advanced Energy Materials* **2020**, *10* (3), 1903176.
(21) Härk, E.; Ballauff, M. Carbonaceous materials investigated by small-angle X-ray and neutron scattering. *C* **2020**, *6* (4), 82.
(22) Lübke, E.; Helfen, L.; Cook, P.; Mirolo, M.; Vinci, V.; Korjus, O.; Fuchsbichler, B.; Koller, S.; Brunner, R.; Drnec, J. The origins of critical deformations in cylindrical silicon based Li-ion batteries. *Energy & Environmental Science* **2024**.
(23) Stevens, D.; Dahn, J. The mechanisms of lithium and sodium insertion in carbon materials. *Journal of The Electrochemical Society* **2001**, *148* (8), A803.
(24) Xie, F.; Xu, Z.; Jensen, A. C.; Ding, F.; Au, H.; Feng, J.; Luo, H.; Qiao, M.; Guo, Z.; Lu, Y. Unveiling the role of hydrothermal carbon dots as anodes in sodium-ion batteries with ultrahigh initial coulombic efficiency. *Journal of Materials Chemistry A* **2019**, *7* (48), 27567-27575.
(25) Härk, E.; Petzold, A.; Goerigk, G.; Ballauff, M.; Kent, B.; Keiderling, U.; Palm, R.; Vaas, I.; Lust, E. The effect of a binder on porosity of the nanoporous RP-20 carbon. A combined study by small angle X-ray and neutron scattering. *Microporous and Mesoporous Materials* **2019**, *275*, 139-146.
(26) Youn, Y.; Gao, B.; Kamiyama, A.; Kubota, K.; Komaba, S.; Tateyama, Y. Nanometer-size Na cluster formation in micropore of hard carbon as origin of higher-capacity Na-ion battery. *npj Computational Materials* **2021**, *7* (1), 48.
(27) CXS group Scanning SAXS Software Package, P., Switzerland, Available from: https://www.psi.ch/en/sls/csaxs/software . (accessed 2023).
(28) Liebi, M.; Georgiadis, M.; Kohlbrecher, J.; Holler, M.; Raabe, J.; Usov, I.; Menzel, A.; Schneider, P.; Bunk, O.; Guizar-Sicairos, M. Small-angle X-ray scattering tensor tomography: Model of the three-dimensional reciprocal-space map, reconstruction algorithm and angular sampling requirements. *Acta Crystallographica Section A: Foundations and Advances* **2018**, *74* (1), 12-24.
(29) Guizar-Sicairos, M.; Thurman, S. T.; Fienup, J. R. Efficient subpixel image registration algorithms. *Optics letters* **2008**, *33* (2), 156-158.

Supporting information

# Resolving the sodiation process in hard carbon anodes with nanostructure specific X-ray imaging

Martina Olsson[1], Antoine Klein[1], Nataliia Mozhzhukhina[1,2], Shizhao Xiong[1], Christian Appel[3], Mads Carlsen[3], Leonard Nielsen[1], Linnea Rensmo[3,4], Marianne Liebi[1,3,4], Aleksandar Matic[1]

*1. Department of Physics, Chalmers University of Technology Gothenburg 41296, Sweden.*

*2. SEEL Swedish Electric Transport Laboratory, Säve Flygplatsväg 27, 42373 Säve, Sweden*

*3. Center for Photon Science, Paul Scherrer Institut, 5232 Villigen PSI, Switzerland*

*4. Institute of Materials, Ecole Polytechnique Fédérale de Lausanne (EPFL), 1015 Lausanne, Switzerland*

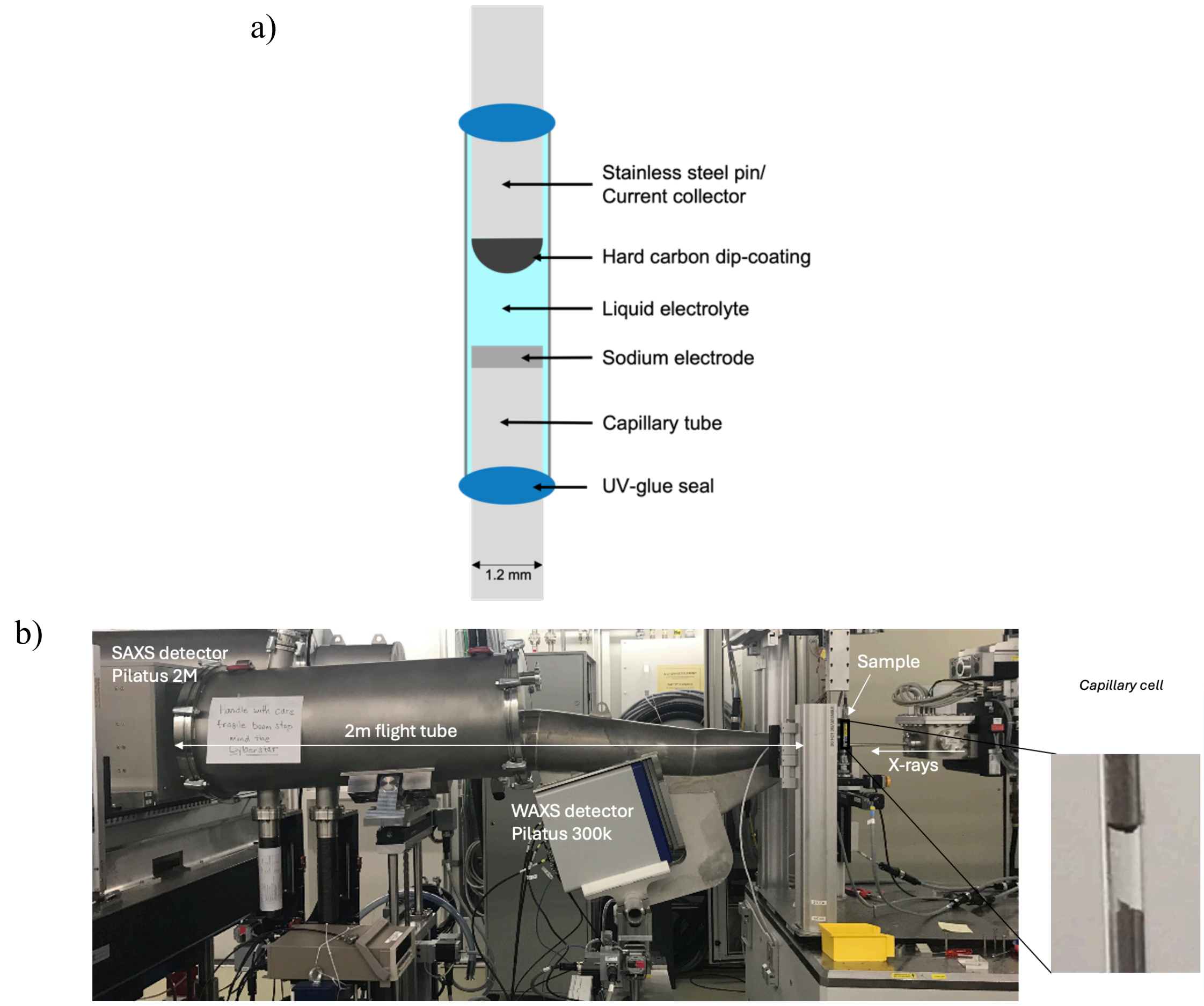

**Figure S1. a)** Schematic image of the capillary-based electrochemical cell used in the experiments**. b)** Experimental set up at cSAXS

**Estimation of charge rate for the coated electrodes**

To estimate the charge rate applied to the different electrodes the volume of the electrodes was calculated from the SAXS imaging results to estimate the mass of each electrode. For the *ex situ* samples the number of voxels in each tomogram of the segmented electrode was calculated while for the operando electrode the volume of the electrode was estimated from the area in the scanning SAXS images by assuming symmetry of the spherical cap. The mass was calculated according to equation S1

$$mass = \#Voxels * Voxel\ dimension * Volume\ fraction\ HC * density\ HC \qquad \text{eq. S1}$$

where the fraction of active hard carbon material is 0.85, the voxel dimension is 15*25*25 $\mu m^3$ and a density of 1.5 $g/cm^3$ of the hard carbon. The resulting C-rate applied to the electrode was further calculated according to equation S2

$$C - rate = \frac{Applied\ current}{Specific\ Capacity * mass} \qquad \text{eq. S2}$$

where a specific capacity of 300 mAh/g was assumed[1]. The result of the calculations for the charge rate of the three electrodes are summarized in Table S1.

**Table S1.** Calculated mass, applied current and the corresponding theoretical C-rate for the different electrodes

| | **Calculated mass** | **Applied current** | **Current density** | **Theoretical charge rate** |
|---|---|---|---|---|
| **Operando cell** | 0.06 mg | 14 µA | 0.23 A/g | 0.8C \| C/1.3 |
| **Low degree of sodiation** | 0.17 mg | 14 µA | 0.082 A/g | 0.3C \| C/3.5 |
| **High degree of sodiation** | 0.22 mg | 14 µA | 0.064 A/g | 0.2C \| C/4.5 |

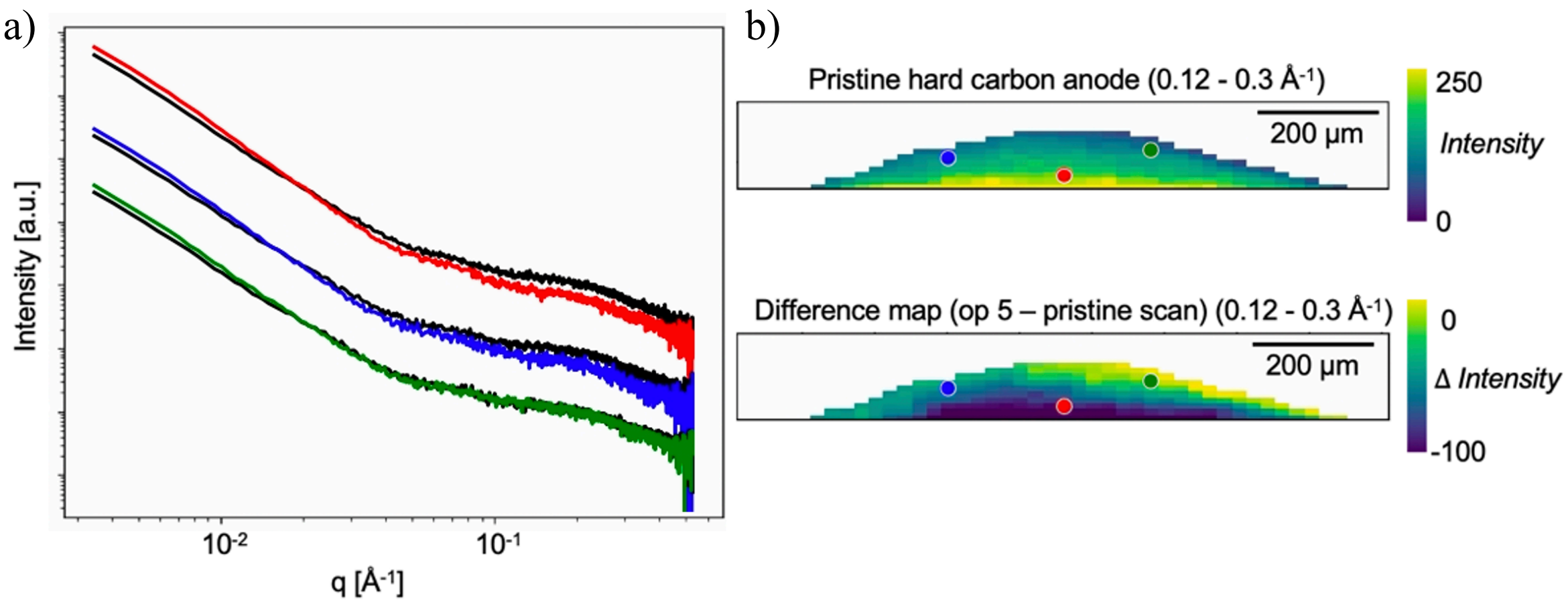

**Figure S2**. Spatially resolved scattering curves at selected points in the pristine and sodiated anode (op 5) **a)** The SAXS curve of the pristine (black) and the sodiated hard carbon (colored) in each point. The coupled curves in each position are offset for clarity. **b)** Scanning SAXS images of the integrated intensity in the microporous (0.12 - 0.3 Å-1) regime of the pristine and sodiated anode (op 5).

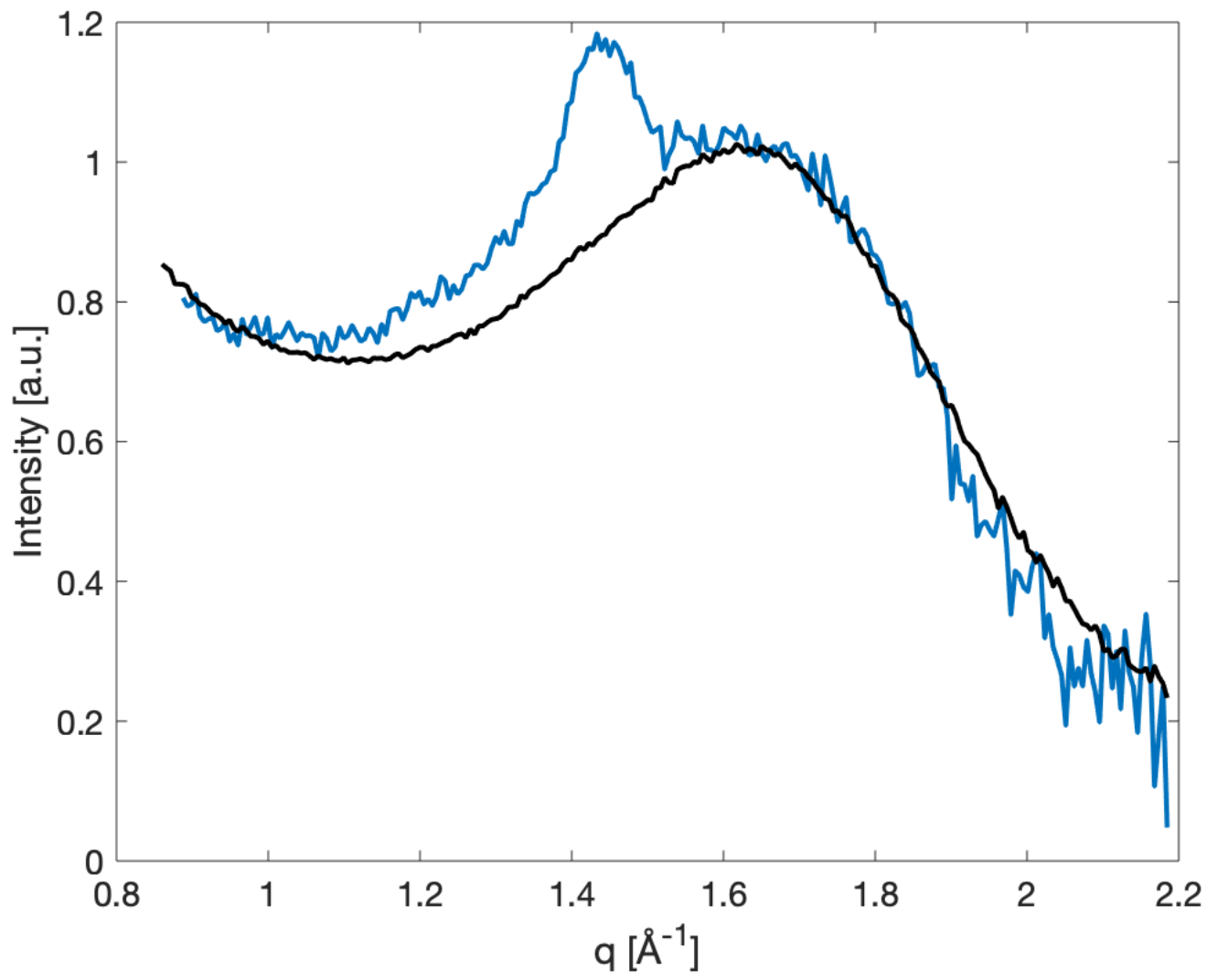

**Figure S3.** WAXS signal from an electrode (blue) including hard carbon as well as the inactive binder material and the pristine hard carbon powder (black). The scattering of the electrode is dominated by the broad 002 peak from the hard carbon and a crystalline peak is observed from the inactive binder material at 1.45 Å$^{-1}$.

**WAXS analysis of the hard carbon electrodes**

To analyze the changes in scattering intensity following sodiation the WAXS curve was divided in different q-regions according to the model contribution described in section 2, equation 4. The q-regions were selected to prevent overlap between different contributions, marked in Figure S4a. Apart from analyzing the scattering intensity changes, the WAXS curves were fitted according to the model described in Figure 4. For the first operando scans a linear background, a gaussian curve describing the 002 peak and a gaussian describing the peak from the inactive binder material within the electrode, (fixed peak position q=1.45 $Å^{-1}$) was used to model the curve. In the last operando scan an additional gaussian was added to account for the broad peak from pseudo-metallic sodium appearing. To calculate the average interlayer distance, $d$, Bragg's law ($d = \frac{2\pi}{q*}$ ) was applied where $q*$ is the 002-peak position. The results are shown in Table S2. Due to the high contribution of electrolyte in the cell which overlaps with the 002-peak the background subtraction impose uncertainties for the extraction of the absolute value of the average interlayer distance and the value should be inferred as a parameter showing the relative change rather than the absolute value of interlayer distance.

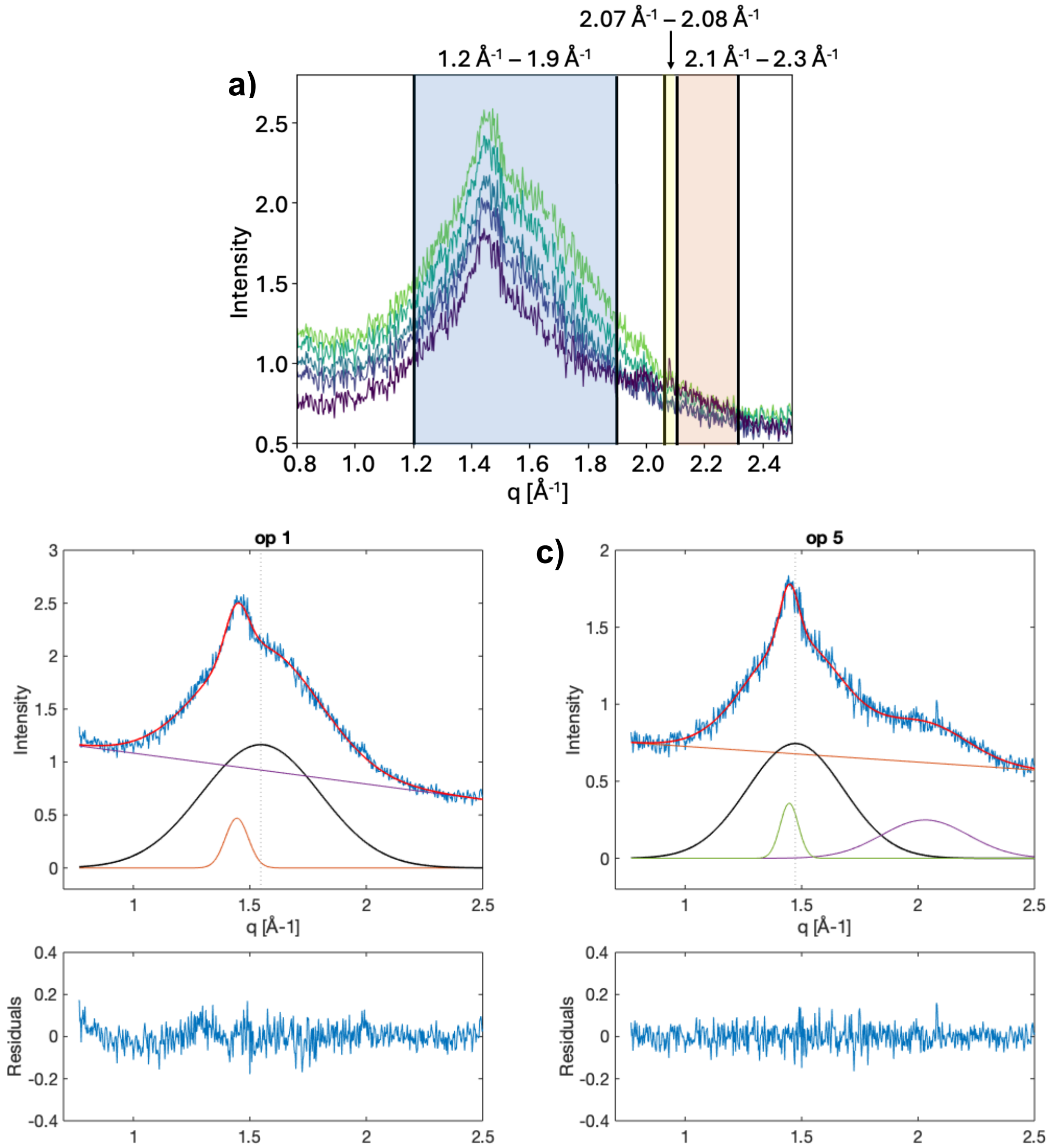

**Figure S4**. **a)** Selected q-regions for intensity analysis in the WAXS region. Fitting results of average WAXS curves from the electrode in the **b)** first operando scan (op 1) and **c)** last operando scan (op 5). The extract 002 peak is shown in black, and the peak position, q*, is marked by a grey dotted line. The fitting was performed with a linear background and a gaussian curve describing the 002 peak as well as a second gaussian describing the peak from the inactive binder material within the electrode, (peak position q=1.45 $Å^{-1}$). In the last operando scan an additional gaussian was added to account for the broad peak of Na-Na correlation from pseudo-metallic sodium confined in the micropores**.**

**Table S2.** Peak position and derived averaged interlayer spacing of the 002-peak in the electrode from the operando WAXS curves. Due to the strong background of electrolyte the values of the average interlayer spacing should be regarded as a relative measurement parameter for comparing the change between scans rather than an absolute measurement.

| | **Peak position 002, q*** | **Average interlayer spacing, 2π/q*** |
|---|---|---|
| **Op 1** | 1.55 $Å^{-1}$ | 4.06 Å |
| **Op 2** | 1.53 $Å^{-1}$ | 4.12 Å |
| **Op 3** | 1.52 $Å^{-1}$ | 4.14 Å |
| **Op 4** | 1.51 $Å^{-1}$ | 4.16 Å |
| **Op 5** | 1.47 $Å^{-1}$ | 4.27 Å |

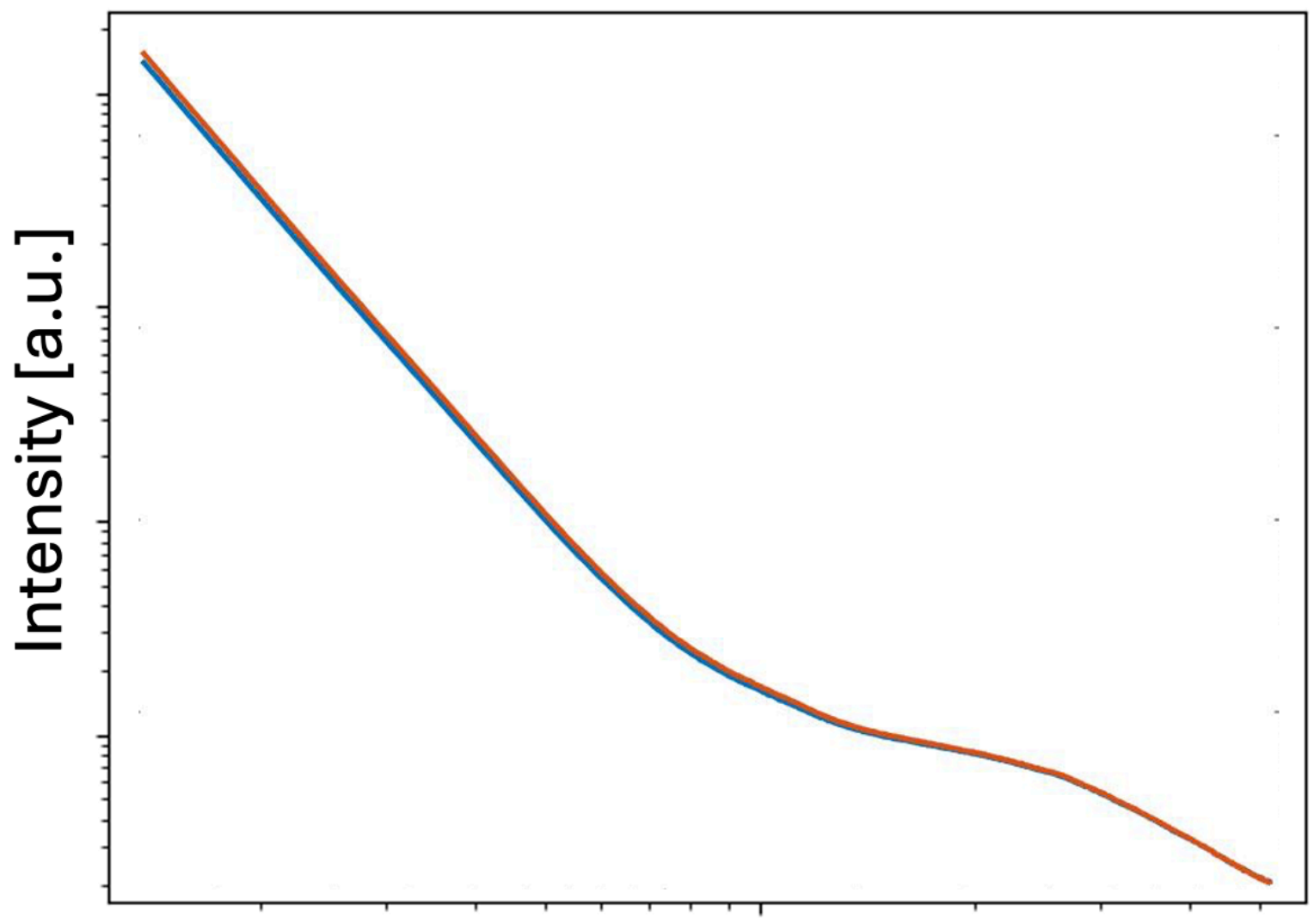

**Figure S5**. Comparison of the average SAXS scattering signal of the first (blue) and last (orange) projection in a tomography measurement of a hard carbon electrode. The curves shows that the radiation dose of the tomogram does not inflict any distinct structural changes and the scattering intensity in the microporous regime remains constant.

(1) Stevens, D.; Dahn, J. High capacity anode materials for rechargeable sodium-ion batteries. *Journal of the Electrochemical Society* **2000**, *147* (4), 1271.